\documentclass[12pt]{article}
\usepackage[centertags]{amsmath}
\usepackage{amsfonts}
\usepackage{amssymb}
\usepackage{amsthm}
\title{\bf Radiation reaction for a massless charged particle}

\author {P. O. Kazinski
and A. A. Sharapov \protect\\
{ \it Physics Faculty, Tomsk State University, Tomsk, 634050  Russia }\\
{\it e-mails:  kpo@phys.tsu.ru,  sharapov@phys.tsu.ru }}

\begin{document}

\maketitle

\begin{abstract}

We derive effective equations of motion for a massless charged
particle coupled to the dynamical electromagnetic field having
regard to the radiation back reaction. It is shown that unlike the
massive case not all the divergences resulting from the
self-action of the particle are Lagrangian, i.e. can be canceled
out by adding appropriate counterterms to the original action.
Besides, the order of renormalized differential equations
governing the effective dynamics turns out to be greater than the
order of the corresponding Lorentz-Dirac equation for a massive
particle. For the case of homogeneous external field the first
radiative correction to the Lorentz equation is explicitly derived
via the reduction of order procedure.

\end{abstract}

\def\thesection{\arabic{section}}
\def\theequation{\arabic{equation}}


\section{Introduction}

One of the interesting features of the classical electrodynamics
is a possibility to account for the self-action of a point charge
in a purely local manner by adding certain higher-derivative terms
to the usual Lorentz's equations. Physically these terms are
responsible for the back reaction of the radiation emitted by an
accelerating charge. The first systematic derivation of the
effective equations of motion for $d=4$ relativistic massive
particle was given by Dirac \cite{Dirac}. Since the
non-relativistic limit of Dirac's equation coincides with the
zero-size limit of Lorentz's model of an electron \cite{Lorentz}
it is often referred to at the Lorentz-Dirac equation. The case of
$d=4$ electrodynamics is not an exceptional one and the analogous
effective equations can be obtained for a massive particle in
arbitrary dimensions, but the local equations take place only for
even $d$'s. The case of $d=6$ massive particle was first
considered in Ref. \cite{Kos}. The general framework to the
Lorentz-Dirac equation in higher dimensions was developed in our
recent paper \cite{KLS}.

Although there is a large literature devoted to  different
aspects of the Lorentz-Dirac equation, its solutions and
applications (see \cite{TVW, {Poisson}, M} for a modern review
and further references) the case of a massless particle has
somehow escaped consideration. The aim of this paper is to fill
this gap in the case of $d=4$ massless particle. We emphasize
that the effective dynamics of  massless particle is essentially
different from that of the massive one, and it can't be obtained
as a massless limit of the Lorentz-Dirac equation.

While massless electrically charged particles (so to speak,
massless electrons or charged photons) have not been yet observed
experimentally no obvious prohibitions are known against their
existence at least from the viewpoint of classical field theory.
Moreover, these particles are predicted by supersymmetric gauge
theories with an unbroken sector of Abelian gauge symmetry.
Perhaps some peculiarities of the effective dynamics, revealed in
this paper, do indicate a certain inconsistency of the classical
electrodynamics of massless particles with radiation effects
included.

This paper may also be viewed as a preparative for study the
radiation reaction problem in the case of extended objects like
strings and branes universally  coupled to the dynamical
antisymmetric tensor fields. It is notable that the
regularization procedure we use here is closely analogous to the
formalism of characteristic (de Rham's) currents intensively
exploited for the study of anomalies in 5-branes coupled to
11-dimensional supergravity background \cite{HL, LM, LMT}.

The paper is organized as follows. In Sec. 2 we discuss some
peculiarities of the dynamics of a relativistic massless particle
coupled to external electromagnetic field and describe a class of
isotropic word lines for which the radiation reaction effects can
be properly treated. Taking into account the radiation back
reaction leads to inevitable infinities resulting from the
''pointness'' of the particle. The renormalization procedure for
these classical infinities is briefly explained in Sec. 3. Here
we also present our main result - the effective equations of
motion for a massless charged particle, and trace the origin of
non-Lagrangian divergences.  As for the Lorentz-Dirac equation,
our equations involve higher derivatives and thus the
reduction-of-order procedure is required to assign them with a
proper mechanical interpretation. This is done in Sec. 4. In
particular, we explicitly derive the leading correction to the
usual Lorentz's equation for the case of homogeneous external
fields. In the concluding section we summarize the results.
Appendix contains details of the calculations omitted in Sec. 3.

\section{Massless particle in the classical electrodynamics}

Let $\mathbb{R}^{3,1}$ be four-dimensional Minkowski space with
coordinates $\{x^\mu\}$, $\mu =0,1,2,3$, and the metric
$\eta_{\mu\nu}=diag(-1,+1,+1,+1)$. Consider a massless point
particle moving in $\mathbb{R}^{3,1}$ and coupled to the
electromagnetic field. The dynamics of the whole system
(field)+(particle) is governed by the usual action
functional\footnote{We take unified dimensions for space and time
setting $c=1$.}
\begin{equation}\label{act}
S=-\frac12\int d\tau e\dot{x}^2 +\int d^4x \left( A_{\mu}j
\,{}^{\mu}-\frac{1}{16\pi} F_{\mu\nu}F^{\mu\nu}\right)\,.
\end{equation}
Here $F_{\mu\nu}=\partial_\mu A_\nu-\partial_\nu A_\mu$ is the
strength tensor of the electromagnetic field, $e\neq 0$ is the
Lagrange multiplier and
\begin{equation} \label{curr}
j^{\mu}(x)=q\int{\delta^4(x-x(\tau))\dot{x}^{\mu}(\tau)d\tau}
\end{equation}
is the density of the electric current produced by the point
charge $q$ moving along a world line $x^\mu(\tau)$ with the
four-velocity $\dot{x}^\mu \equiv dx^\mu/d\tau$. The action
(\ref{act}) is invariant under  arbitrary world-line
reparametrizations $\tau\rightarrow\tau'$, provided $e(\tau)$ is
transformed as a world-line density (einbein)
\begin{equation} \label{tp}
e(\tau')= \frac{d\tau'}{d\tau}e(\tau)\,.
\end{equation}
Varying the action (\ref{act}) with respect to $e$ one gets the
standard isotropy condition (Lagrangian constraint) on the
four-velocity of the massless particle
\begin{equation}\label{is}
\dot{x}^2=0\,,
\end{equation}
i.e. a tangent vector to each point of particle's trajectory
$x^\mu(\tau)$ lies on the future light cone attached at this
point. We will refer to such a trajectory as an isotropic one. The
variations of the action with respect to particle's trajectory
$x^\mu(\tau)$ and electromagnetic potentials $A_\mu(x)$ lead to
the coupled systems of the Lorentz and Maxwell equations
\begin{equation} \label{leq}
e\ddot{x}_\mu+\dot{e}\dot{x}_\mu=qF_{\mu\nu}\dot{x}^{\nu},
\end{equation}
\begin{equation} \label{meq}
\partial^{\nu}F_{\mu\nu}=4\pi{}j_{\mu}\,.
\end{equation}
Using the reparametrization invariance of the model and the
transformation property (\ref{tp}) one can always bring the
einbein $e$ into the form $e=1$ or, equivalently, one may impose
this condition to fix a particular parametrization of the world
line. In this gauge the form of the Lorentz equations is quite
similar to that for a  massive particle,
\begin{equation} \label{feqm}
\ddot{x}_\mu=qF_{\mu\nu}\dot{x}^{\nu}\,,
\end{equation}
written in a \textit{proper-time parametrization} ($\dot{x}^2=1$).
The crucial difference, however, is a presence of the additional
isotropy condition (\ref{is}) having no analog in the massive
case. In particular this suggests impossibility to introduce the
proper-time parametrization for world lines of the massless
particle.

Another convenient  parametrization to use when working with the
isotropic world lines is extracted by the following gauge fixing
condition:
\begin{equation}\label{gfc}
\ddot{x}^2=1\,.
\end{equation}
One may regard it as a ``massless'' counterpart of the
proper-time parametrization. Contrary to the massive case,
however, there are some isotropic trajectories which do not admit
such a parametrization. Since for any isotropic curve
$\ddot{x}^2\geq 0$, the trajectories for which $\ddot{x}^2=0$ at
some  $\tau$'s may be thought of as \textit{degenerate} ones in
the sense that the complementary set of nondegenerate trajectories
fills an open and everywhere dense domain in the space of all
isotropic curves, so that any degenerate isotropic curve can be
made nondegenerate by a small perturbation.

On the other hand, if for some interval $\tau_1<\tau<\tau_2$ the
four-vector of acceleration is know to be isotropic,
$\ddot{x}^2=0$, then we can immediately conclude that this part of
the trajectory is given by a straight segment. Indeed, since the
velocity and acceleration of the particle are always orthogonal to
each other we get a pair of isotropic and orthogonal vectors
$\dot{x}^\mu$ and $\ddot{x}^\mu$, but any such two vectors are
known to be proportional to each other in the Minkowski space,
\begin{equation}\label{}
  \ddot{x}^\mu=\lambda(\tau)\dot{x}^\mu\,.
\end{equation}
Integrating the last equation we see that the particle does move
along a straight line until $\tau_1<\tau <\tau_2$.

The positive-curvature condition ($\ddot{x}^2>0$) for particle's
trajectory implies a nonzero value of the external electromagnetic
field in the right-hand side of the Lorentz equation (\ref{leq}),
otherwise the particle would move along a straight isotropic line
for which $\ddot{x}^\mu=0$. More precisely, squaring both sides
of the Lorentz equation and accounting that, by definition,
$e\neq 0$ we arrive at the following condition:
\begin{equation}\label{cc}
  T_{\mu\nu}(x)\dot{x}^\mu\dot{x}^\nu > 0\,,
\end{equation}
where
\begin{equation}\label{}
  T_{\mu\nu}=-\frac{1}{4\pi} \left( F_\mu{}^\alpha F_{\alpha\nu} +\frac14\eta_{\mu\nu}
  F^{\alpha\beta}F_{\alpha\beta} \right )
\end{equation}
is the stress-energy tensor of the electromagnetic field. Passing
to an appropriate Lorentz frame it is not hard to see that
$T_{\mu\nu}\dot{x}^\mu\dot{x}^\nu\geq 0$ for any field
$F_{\mu\nu}=(\mathbf{E},\mathbf{H})$, and the equality implies
either
\begin{equation}\label{}
 \mathbf{v}\parallel \mathbf{E}\parallel\mathbf{H}
\end{equation}
or
\begin{equation}\label{}
  \mathbf{v}\parallel \mathbf{E}\times\mathbf{H}\,\;\;\;\;
  {\rm and}\;\;\;\;\mathbf{E}\cdot \mathbf{H}=0\,,\;\;\mathbf{E}^2=\mathbf{H}^2\,,
\end{equation}
$\mathbf{v}=\{\dot{x}^i\}$ being the space part of the
four-velocity. The latter situation may occur, for example, for a
massless particle moving on a background of free electromagnetic
wave.

In this paper we restrict our attention to the self-consistent
solutions for the Lorentz-Maxwell equations (\ref{leq}, \ref{meq})
obeying the gauge fixing condition (\ref{gfc}) and the
compatibility condition (\ref{cc}). In the next section we will
see how these restrictions arise naturally upon calculating the
radiation-reaction force.

\section{Radiation reaction and renormalization}

In this section we follow the approach of work  \cite{KLS} to
account for the radiation back reaction to the relativistic
motion of a point charge. For this end, we first solve the
Maxwell equations (\ref{meq}) expressing the electromagnetic
field as a functional of particle's trajectory. The effective
equations of motion are then obtained by substituting this
solution to the Lorentz equations (\ref{leq}). Inasmuch as the
electromagnetic field is singular at the points of particle's
trajectory the last step requires a renormalization procedure to
remove inevitable infinities. The alternative approach to the
problem (also involving some renormalization) is based on the
computation of energy radiated by an accelerating charge and
identification of this energy with the work of a hypothetical
(radiation reaction) force \cite{Dirac, Kos, TVW, Poisson}.

Let us proceed to the calculations. In the Lorentz gauge
$\partial^\mu A_\mu=0$ the Maxwell equations take the form
\begin{equation}\label{}
\square A_\mu(x)=-4\pi j_\mu(x)\,.
\end{equation}
Any solution to these equations may be constructed as the sum of
their particular solution and a solution to the corresponding
homogeneous equations. Although this decomposition is quite
ambiguous (from the pure mathematical viewpoint) it has a certain
physical meaning if one identifies the former solution with the
own field of the point charged particle,  treating  the latter as
an external field describing free electromagnetic waves incident
on the particle. The own field of the particle is given by the
Li\'enard-Wiechert potentials
\begin{equation} \label{pot}
A^{LW}_{\mu}(x)=-4\pi{}\int{G^{ret}(x-z)j_{\mu}(z)d^4z}\,,
\end{equation}
where
\begin{equation}\label{gf}
G^{ret}(x)=-\frac{1}{2\pi}\theta (x^0 )\delta (x^2)
\end{equation}
is the retarded Green's function. Taking together Rels.
(\ref{curr}), (\ref{pot}), (\ref{gf})  we arrive at the standard
expression for the Lorentz force
$F_\mu(s)=F_{\mu\nu}(x(s))\dot{x}^{\nu}(s)$ accounting for the
self-action of the particle
\begin{equation}\label{force}
\begin{array}{c}
F_\mu=F_\mu^{ext}+F_\mu^{rr}\,,\;\;\;\;\;\;\;
F^{ext}_\mu(s)=F_{\mu\nu}^{ext}(x(s))\dot{x}^\nu(s)\,,\\[5mm]
\displaystyle{F^{rr}_\mu (s) =
F_{\mu\nu}^{LW}(x(s))\dot{x}^\nu(s)=4q^2\int\limits_{-\infty}^s
\delta '(-X^2(s,\tau ))\dot x_{[\mu } (\tau )X_{\nu ]}(s,\tau)
\dot
x^\nu (s)d\tau}\,,\\[8mm]
X^\mu(s,\tau)\equiv x^\mu(s)-x^\mu({\tau})\,,
\end{array}
\end{equation}
where $F^{ext}_{\mu\nu}$ is the strength of the external
electromagnetic field and the square brackets stand for the
antisymmetrization of indices (without one-half). The
abbreviation $``rr"$ labeling the self-action part of the Lorentz
force points on the common interpretation of this term as that
describing the \textit{radiation reaction}, i.e. the back reaction
of the radiation emitted by an accelerating charge.

Note that expression (\ref{force}) is not very meaningful as it
stands since the integral diverges. The divergence arise from the
singularity of Green's function (\ref{gf}) at the vertex of the
future light cone $x^2=0$, $x^0>0$. So we need some
regularization procedure smoothing the behaviour of Green's
function $G(x)$ at $x=0$. The most simple and efficient way to do
this is to replace the $\delta$-function entering the expression
for $G$ by the following $\delta$-shaped sequence \cite{KLS}:
\begin{equation}\label{reg}
\delta(s)=\lim_{a\rightarrow +0}\frac{e^{-s/a}}{a}\,,\;\;\;\;s\geq
0\,.
\end{equation}
Integrating this expression  with a test function on the positive
half-line one can obtain the following asymptotic expansion:
\begin{equation} \delta_a(s)=\frac{e^{-
s/a}}{a}=\sum_{n=0}^{\infty}(-a)^n\delta^{(n)}(s)\,.
\end{equation}
Thus, the regularized expression for the radiation reaction force
(\ref{force}) can be written as
\begin{equation}\label{flornonreg}
F^{rr}_\mu(s,a)=-4q^2\dot{x}^\nu (s)\frac \partial {\partial
a}\int\limits_0^\infty \frac{d\tau}{a}\;{ e^{X^2(s,s-\tau)/a}}
X_{[\nu}(s,s-\tau)\dot{x}_{\mu] }(s-\tau ) \,,
\end{equation}
\begin{equation*}
F^{rr}_\mu (s)=\lim_{a\rightarrow +0}F^{rr}_\mu(s,a)\,.
\end{equation*}
This integral is localized at the point $\tau=0$ as $a\rightarrow
+0$, and hence, can be evaluated by the Laplace method \cite{GS}.
In a small vicinity of the localization point the damping
exponential factor behaves as
\begin{equation}\label{exp}
e^{X^2(s,s-\tau)/a}=\exp\left(\dot{x}^2\frac{\tau^2}a-\dot
x\cdot\ddot x \frac {\tau ^3}a+\left[\frac 14\ddot{x}^2+\frac13
\dot x\cdot\dddot x \right]\frac{\tau ^4}a+O(\tau^5) \right)\,.
\end{equation}
Note that for the massive particle  $\dot{x}^2 < 0$ and the
asymptotic of the Laplace integral (\ref{flornonreg}) is governed
by the leading $\tau^2$-term, while for the massless particle it
is determined by $\tau^4$-term. In the latter case
\begin{equation}
\dot{x}^2=\dot{x}\cdot\ddot{x}=0\,,\;\;\;\;\;\;
\frac14\ddot{x}^2+\frac13\dot{x}\cdot\dddot{x} =
-\frac{1}{12}\ddot{x}^2\,,
\end{equation}
and the coefficient at $\tau^4$ is strongly negative if we demand
$\ddot{x}^2=1$. In such a way we recover the positive-curvature
condition for the isotropic world lines discussed in the previous
section. In the massless case the result of integration should
have a form of Laurent's series in $a^{1/4}$ with a finite
irregular part. The details of calculations are given in
Appendix. The result is
\begin{equation}\label{rrf}
\begin{array}{c}
\displaystyle{    F^{rr}_\mu(s,a) =
-q^2\left[\frac{\alpha_1}{4}\ddot{x}_{\mu}a^{-3/4}+
    \frac{\alpha_2}{10}\dddot{x}^2\dot{x}_{\mu}a^{-1/2}-
    \frac{\alpha_3}{16}\left(\stackrel{(4)}{x}_\mu+\frac{11}{10}\dddot{x}^2\ddot{x}_\mu+
    \frac{11}{5}\dddot{x}\cdot\stackrel{(4)}{x}\dot{x}_\mu\right)a^{-1/4}\right]-}\\[8mm]
\displaystyle{
-{q^2}\frac25\left\{\stackrel{(5)}{x}_\mu+{\dddot{x}^2}\dddot{x}_\mu+
    3{\dddot{x}\cdot\stackrel{(4)}{x}}\ddot{x}_\mu+\left(\frac{9}{7}\stackrel{(4)}{x}{}^2+
    \frac{11}{7}\dddot{x}\cdot\stackrel{(5)}{x}+\frac{2}{5}(\dddot{x}^2)^2\right)\dot{x}_\mu\right\} }+\cdots
    ,\\[8mm]
    \alpha_n=12^{n/4}\Gamma\left(1+n/4\right)\,,
\end{array}
\end{equation}
where dots stand for terms vanishing upon removing regularization.
The expression (\ref{rrf}) contains three singular terms, two of
which are Lagrangian, namely, those at $-3/4$ and $-1/4$ powers
of $a$, while the rest is not. For example, $a^{-3/4}$-term is
obtained by varying the following functional:
\begin{equation}\label{}
q^2\frac{\alpha_1}{8}a^{-\frac34}\int{\frac{\dot{x}^2}{\sqrt[4]{\ddot{x}^2}}d\tau}
\,.
\end{equation}
As to $a^{-1/4}$-term, the corresponding Lagrangian is given by a
rather unwieldy expression. In order to simplify it let us
redefine the Lagrangian multiplier $e$ entering the l.h.s. of the
Lorentz equation (\ref{leq}) as follows:
\begin{equation}\label{ree}
    e\rightarrow e-q^2\frac{11}{160}\alpha_3\dddot{x}^2a^{-1/4}\,.
\end{equation}
Then the expression for $a^{-1/4}$-term reduces to
\begin{equation}\label{reterm}
   q^2 \frac{\alpha_3}{16}\stackrel{(4)}{x}_\mu a^{-1/4}\,,
\end{equation}
which can be obtained by varying the following functional:
\begin{equation}\label{}
    q^2\frac{\alpha_3}8a^{-1/4}\int{\sqrt[4]{\ddot{x}^2}}d\tau
\end{equation}

A practical recipe to restore both the Lagrangians is as follows.
We exclude the electromagnetic field $A_\mu(x)$ from the action
functional (\ref{act}) by the Maxwell equations (\ref{meq}). The
result is the Fokker type \cite{Fokker} effective action for the
charged massless particle
\begin{equation}\label{effact}
\begin{array}{r}
\displaystyle S_{eff}=-\frac12\int d\tau e\dot x^2-2\pi\int
d^4x\int
d^4yj^\mu (x)G^{ret}(x-y)j_\mu (y)=  \\[5mm]
\displaystyle =-\frac12\int d\tau e\dot x^2+\frac{q^2}{2}\int
ds\int d\tau \;\dot x^\mu (s)\delta(X^2(s,\tau))\dot x_\mu (\tau
)\,.
\end{array}
\end{equation}
The first term is the usual action of a free massless particle,
while the second term describes particle's self-action. Using
regularization (\ref{reg}) for the $\delta$-function entering the
self-action term we may perform one of two integrations. The
calculations are quite similar to those presented in  Appendix.
The only distinction is that now we cannot use the isotropy
condition (\ref{is}). Instead, we rewrite the exponent
(\ref{exp}) in the following way:
\begin{equation}\label{}
\exp\left(\left[\frac 14\ddot{x}^2+\frac13 \dot x\cdot\dddot x
\right]\frac{\tau ^4}a\right)\exp\left(
\dot{x}^2\frac{\tau^2}a-\dot x\cdot\ddot x \frac {\tau
^3}a+O(\tau^5)\right)\,.
\end{equation}
Then the first multiplier is interpreted as a damping exponential
factor of the Laplace integral, while the second is combined with
the rest of the integrand and expanded in the Taylor series in
$\tau$; in so doing, we may keep only terms which are at most
linear in $\dot x^2$ and $\dot x\cdot \ddot x$, since the higher
powers will not contribute to the equations of motion $\delta
S_{eff}/\delta x=0$ (more precisely, their contribution will
vanish on the shell of the isotropy condition $\dot x^2=\delta
S_{eff}/\delta e=0$). Integrating we get Laurent's series in
$a^{1/4}$ without a constant term and with two singular terms
giving upon variation two of three foregoing singularities.
Clearly, the Lagrangians obtained in such a way are not uniquely
determined as one may add to them any local expression
proportional to the squares of the Lagrangian constraint $\dot
x^2$ and its differential consequences.

The reason why performing a self-consistent elimination of the
electromagnetic field from the quadratic (in $A_\mu$) action
(\ref{act}) we do not reproduce the whole expression for the
radiation reaction force (\ref{rrf}) including the finite part
and the $a^{-1/2}$-term is as follows: Only symmetric part
\begin{equation}\label{}
G^{ret}(x-y)+G^{ret}(y-x)=-\delta((x-y)^2)
\end{equation}
of the retarded Green's function (\ref{gf}) actually contributes
to the effective action (\ref{effact}), while substituting the
Li\'enard-Wiechert potentials to the Lorentz equation (\ref{leq})
we use, in fact, the entire Green's function. The symmetric and
antisymmetric parts of the Green function lead to the different
groups of terms in (\ref{rrf}); the first terms are invariant
under reversion of particle's trajectory $\tau\rightarrow -\tau$,
whereas the second ones acquire the minus sign. Therefore only
the invariant terms can arise from the effective action
(\ref{effact}).

According  to general prescriptions of the renormalization theory
the singular coefficients $a^{-3/4}$, $a^{-1/2}$, $a^{-1/4}$ in
the regularized expression (\ref{rrf}) are replaced by finite
constants to be fixed from an experiment. Since there is no
evidence for a preferable choice for these constants we may put
them to zero to simplify further consideration. Then the
radiation reaction force $F_\mu^{rr}$ is given by the second line
in Eq.(\ref{rrf}). The non-Lagrangian nature of this force
manifests itself in the violation of the time-reversion symmetry,
$\tau \rightarrow -\tau$, which is consistent with
irreversibility of the radiation process.

In conclusion of this section let us note that although the
expression (\ref{rrf}) for the radiation reaction force was
derived in a special gauge, it is not hard to rewrite it
 in an arbitrary parametrization. For  this end one should just treat the
overdot derivatives as invariant ones,
\begin{equation}\label{}
D= \frac{1}{\sqrt[4]{\ddot{x}^2}}\frac{d}{d\tau}\,,
\end{equation}
where $\tau$ is already an arbitrary evolution parameter. Note
that in an arbitrary parametrization the effective equations of
motion for a charged massless particle involve a sixth derivative
(for comparison, the Lorentz-Dirac equation for d=4 massive
particle is of the third order).

\section{Reduction of order \protect \footnote{We are thankful
to S.L. Lyakhovich for illuminating discussions on various aspects
of the reduction-of-order procedure for higher-derivative
Lagrangian systems.} }

Let us write down the effective equation of motion governing the
dynamics of a massless charged particle,
\begin{equation}\label{eq}
\begin{array}{l}
\displaystyle{ D(\tilde{e} Dx ^\mu)=\Omega^{\mu}_{\nu} Dx^{\nu} -
 \frac {2q}{5}  \left(D^5x^\mu+(D^3x)^2D^3x^\mu+3D^3x\cdot D^4 x
D^2x^\mu+\right.}\\[5mm]
\displaystyle{\left. +\left[\frac97(D^4x)^2+
 \frac{11}7D^3x\cdot D^5x+\frac25((D^3x)^2)^2  \right]Dx^\mu
 \right)}\,,\;\;\;\;\;\;\;(Dx)^2=0\,,\\[5mm]
\end{array}
\end{equation}
where $\tilde{e}(\tau)=q^{-1}e(\tau)\sqrt[4]{\ddot{x}^2}$ is a
nonzero  scalar function and
$\Omega^\mu_\nu=\eta^{\mu\lambda}F^{ext}_{\lambda\nu}(x)$.

Since the order of the equation is greater than two, it cannot be
assigned with a straightforward mechanical interpretation: In the
realm of Newtonian mechanics a state of the particle is
unambiguously determined by specifying  its position and
velocity, that is obviously insufficient to extract a particular
solution to our equation. This problem is analogous to that of a
mechanical interpretation of the Lorentz-Dirac equation which also
has too many solutions, but not all of them are physically
meaningful. To extract the subspace of physical solutions we
impose an additional selection rule \cite{Bh}: The physical
solutions to the Eq. (\ref{eq}) are only those which have a
smooth limit upon switching off the interaction, i.e. when $q
\rightarrow 0 $. In other words, we treat the radiation reaction
as a small perturbation deforming the conventional dynamics of
the charged particle (i.e. the dynamics without account of the
self-action) rather than something introducing extra degrees of
freedom to the theory. This treatment is justified by a small
parameter $q$ at the higher-derivative terms in the r.h.s. of the
Eq.(\ref{eq}). Instead of seeking for solutions which can be
smoothly continued to $q=0$, one may equivalently seek for a
second-order differential equation (with a  smooth dependence of
$q$) any solution of which would be a solution to the initial
equation. The general procedure for obtaining such an equation is
known as the \textit{reduction of order}. It consists in
successive elimination of the higher derivatives from the
equation (\ref{eq}) with the help of all its differential
consequences; in so doing, the order of the equation increases at
each step of the procedure, but the order of $q$ at higher
derivatives increases as well, so that, in the limit, we get a
power series in $q$ with coefficients depending on $x, Dx$ and
$D^2x$. The implementation of this procedure to the Lorentz-Dirac
equation may be found, for example, in \cite{LL, Poisson, M}. It
should be noted, however, that the resulting equation for the
massless particle appears to be unresolved with respect to the
highest (second) derivative that poses the questions about
existence and uniqueness of its solution given the initial
position and velocity of the particle. Without going into detail
we just note that for a homogeneous external field
($\Omega=const$) the reduced equation can be perturbatively
resolved with respect to $D^2x$ with the right hand side given by
a power series in $q$. Up to the first order in $q$ it reads
\footnote{Here we use matrix notation: $(\Omega\dot
x)^\mu=\Omega^\mu_{\nu}\dot x^\nu$, $\;\dot x\Omega^2\dot x=\dot
x_\nu \Omega ^\nu_\sigma \Omega^\sigma_\mu \dot x^\mu$, and so
on.}
\begin{equation}\label{reduced}
\begin{array}{c}
\displaystyle{\ddot{x}^\mu=-\tilde{e}^{-1}(\Omega \dot{x})^\mu+}\\[5mm]
\displaystyle{+\frac{2q}5
\left(\frac{(\Omega^4\dot{x})^\mu}{\tilde{e}^5}+\frac{\dot{x}\,\Omega^4\dot{x}}{
\tilde{e}^7}(\Omega^2\dot{x})^\mu
+\frac12\left[\frac97\frac{\dot{x}\,\Omega^6\dot{x}}{\tilde{e}^7}+
\frac75\frac{(\dot{x}\,\Omega^4\dot{x})^2}{\tilde{e}^9}\right]\dot{x}^\mu
\right)+O(q^2)}\,,
\end{array}
\end{equation}
where overdot stands for the invariant derivative $D$ and the
function  $\tilde{e}$, being determined from the identity
$\ddot{x}^2=1$, is given by
\begin{equation}\label{}
  \tilde{e}^2=-\dot{x}\Omega^2\dot{x}+O(q^2)\,.
\end{equation}
As above we are restricted to initial velocities obeying condition
$\tilde{e}\neq  0$ (compare with (\ref{cc})) which ensures
existence of a solution for a sufficiently small time interval. In
the gauge $e=1$ the above equation takes the form
\begin{equation}\label{reduced1}
\begin{array}{c}
\displaystyle{\ddot{x}^\mu=q(\Omega \dot{x})^\mu+}\\[5mm]
\displaystyle{+\frac{2q^4}5
\left(\frac{(\Omega^4\dot{x})^\mu}{\dot x\Omega^2\dot x
}-\frac{\dot{x}\,\Omega^4\dot{x}}{ (\dot x\Omega^2\dot x
)^2}(\Omega^2\dot{x})^\mu
-\frac12\left[\frac97\frac{\dot{x}\,\Omega^6\dot{x}}{(\dot
x\Omega^2\dot x )^2}-
\frac75\frac{(\dot{x}\,\Omega^4\dot{x})^2}{(\dot x\Omega^2\dot x
)^3}\right]\dot{x}^\mu \right)+o(q^4)}\,,
\end{array}
\end{equation}
where overdot denotes now the ordinary derivative w.r.t. the
evolution parameter. Note that although the condition
 $(\dot x\Omega^2\dot x)<0$ has been assumed at each stage of our
derivation the resulting equation has the correct
\textit{uniform} limit upon switching off the external
electromagnetic field (i.e. if we set $\Omega=\varepsilon\Omega'$
and then let $\varepsilon\rightarrow 0$). The last statement
remains true even with account of higher orders in $q$, so the
limiting equation $\ddot x^\mu =0$ describes the free motion.

Thus, the perturbative  treatment of the radiation reaction,
being applied to the case at hands, leads to the meaningful second
order equation at least for the constant fields.

\section{Conclusion}

To summaries, in this paper we have derived the effective
equations of motion for a massless charged particle interacting
with external electromagnetic field as well as its own one.
Although the approach we follow here is quite similar to that
used in the case of a massive particle \cite{KLS} the final
results considerably differ from each other. First, not all the
divergencies are Lagrangian, i.e. may be canceled out by adding
appropriate counterterms to the original action functional; even
though such counterterms do exist (for some divergences) they are
not uniquely defined. The ambiguity reflects the freedom to add to
the Lagrangian any local expression proportional to the square of
the isotropy condition and its differential consequences. The
non-Lagrangian divergences in turn are recognized as those
breaking up the time-reversion symmetry. Second, consistency of
the dynamics imposes some restrictions on initial data of the
particle and/or the form of external fields. Third, after
renormalization we get a fifth order differential equation,
instead of third-order equation derived by Dirac for a massive
particle.

All these distinctions have a common origin related to the special
local structure of the isotropic world lines: The interval
between two nearby points $x(\tau)$ and $x(\tau+\delta\tau)$ on
an isotropic curve, lying in a general position, is proportional
to $\delta\tau^2$, rather than to be linear in $\delta \tau$ as is
case for the time-like world lines of massive particles. This
leads to the stronger singularities, as compared to the massive
case, and as result to a different structure of the radiation
reaction force.

Finally, the straightforward application of the
reduction-of-order procedure leads to the differential equation in
a form unresolved with respect to the second derivative which may
lead to some difficulties upon its integration. Besides, it may
have no well-defined limit upon switching off an external
electromagnetic field as seen from the structure of the first
radiative correction (\ref{reduced1}). The last fact may indicate
a certain instability of the classical dynamics and deserve a
further study.

It would be interesting to re-derive our equation on the basis of
usual energy conservation arguments in order to gain a more
physical insight into the problem as well as an independent test
for consistency of the formal renormalization procedure we have
used.

\section*{Appendix}

The regularized expression for the self-action force
(\ref{flornonreg}) involves the following integral:
\begin{equation}\label{I}
I_{\mu}(s,a)=\dot{x}^\nu(s)\frac{\partial}{\partial
a}\int\limits_0^\infty \frac{d\tau}{a}\;
X_{[\nu}(s,s-\tau)\dot{x}_{\mu] }(s-\tau ){ e^{X^2(s,s-\tau)/a}}
\,,
\end{equation}
Here we apply the Laplace method to obtain the Laurent series for
the function $I_{\mu}$ with respect to $a^{1/4}$. More precisely,
we are interested only in  constant and  irregular
 parts of this series (contributing, respectively, to the
finite and the divergent parts the self-action force) since the
regular terms vanish upon switching off the regularization.

In order to simplify formulae we use standard notation from the
linear algebra for the inner and the exterior (skew-symmetric)
products of two vectors: $a\cdot b=a^\mu b_\mu$, $a\wedge b=(a_\mu
b_\nu-a_\nu b_\mu)$. Thus, the pre-exponential  factor in the
integral (\ref{I}) may be written as
\begin{equation}\label{b}
\dot x(s-\tau)\wedge X(s,s-\tau)=\sum_{m=2}^{\infty}\tau^m
b_m(s)\,,
\end{equation}
where
$$
b_m=(-1)^m\sum_{n=1}^m\frac{\stackrel{(n)}{x}\wedge \stackrel{(m-n+1)}{x}}{%
n!(m-n)!}\,.
$$
In particular,
$$
b_2=-\frac12 \stackrel{(2)}{x}\wedge \stackrel{(1)}{x}\,,\;\;\;\;
b_3=\frac 13\stackrel{(3)}{x}\wedge \stackrel{(1)}{x}\,,
$$

$$
b_4=-\frac 18\stackrel{(4)}{x}\wedge
\stackrel{(1)}{x}-\frac 1{12}\stackrel{(3)}{x}\wedge
\stackrel{(2)}{x}\,,\;\;\;\; b_5=\frac
1{30}\stackrel{(5)}{x}\wedge \stackrel{(1)}{x}+\frac 1{24}
\stackrel{(4)}{x}\wedge \stackrel{(2)}{x}\,,
$$

$$
b_6=-\frac 1{144}\stackrel{(6)}{x}\wedge \stackrel{(1)}{x}-\frac
1{80} \stackrel{(5)}{x}\wedge \stackrel{(2)}{x}-\frac
1{144}\stackrel{(4)}{x} \wedge \stackrel{(3)}{x}\,,\;\;\;
b_7=\frac 1{840}\stackrel{(7)}{x}\wedge \stackrel{(1)}{x}+\frac
1{360} \stackrel{(6)}{x}\wedge \stackrel{(2)}{x}+\frac
1{360}\stackrel{(5)}{x} \wedge \stackrel{(3)}{x}\,.
$$
Using the basic identities
$$
\dot x^2=0\,,\;\;\;\;\;\;\; \ddot x^2 =1
$$
as well as their differential consequences
$$
\begin{array}{llll}
\ddot x \cdot \dot x=0\,, & \dddot x\cdot\ddot x=0\,,&\dddot
x\cdot\dot
x=-1\,, & \stackrel{(4)}{x}\cdot \dot x=0\,,\\[5mm]
\stackrel{(4)}{x}\cdot \stackrel{(2)}{x}+\stackrel{(3)}{x}{}^2\,=&
\stackrel{(5)}{x}\cdot
\stackrel{(1)}{x}-\stackrel{(3)}{x}{}^2\,=&\stackrel{(5)}{x}\cdot
\stackrel{(2)}{x}+3\stackrel{(4)}{x}\cdot\stackrel{(3)}{x}\,
=&\stackrel{(6)}{x} \cdot
\stackrel{(1)}{x}-5\stackrel{(4)}{x}\cdot \stackrel{(3)}{x}=0\,,
\end{array}
$$
the argument of the exponent (\ref{I}) can be written as
$$
X^2(s,s-\tau)=(x(s-\tau)-x(s))^2=\sum_{k=4}^\infty  c_k\tau^k=
$$
\begin{equation}\label{X}
=-\frac 1{12}\stackrel{(2)}{x}\cdot \stackrel{(2)}{x}\tau^4+\frac
1{360}\stackrel{(3)}{x}\cdot \stackrel{(3)}{x}\tau^6-
\end{equation}

$$
-\frac 1{360} \stackrel{(3)}{x}\cdot
\stackrel{(4)}{x}\tau^7+\left(\frac 1{1260}\stackrel{(3)}{x}\cdot
\stackrel{(5)}{x} +\frac 1{1344}\stackrel{(4)}{x}\cdot
\stackrel{(4)}{x}\right)\tau^8+o(\tau^8)\,,
$$
$$ $$
where
$$
c_k=(-1)^k\sum_{n=1}^{k-1}\frac{\stackrel{(n)}{x}\cdot
\stackrel{(k-n)}{x}}{n!(k-n)!}\,.
$$
Substituting (\ref{b}) and (\ref{X}) to the integral (\ref{I}) and
making replacement $t\rightarrow a^{1/4}t$ we find that
\begin{equation}
I_\mu (a)=\dot x^\nu \frac{\partial}{\partial a}\int_0^{\infty}
d\tau\left(\sum_{m=3}^7a^{(m-3)/4}\tau^m
(b_m)_{\nu\mu}\right)\exp\left(
\sum_{n=4}^8a^{(n-4)/4}\tau^nc_n\right)+ {\rm (regular\;terms)}
\end{equation}
The expansion for the exponent reads
\begin{equation}
e^{-\tau^4/12}(1+a^{1/2}\tau^6c_6+a^{3/4}\tau^7c_7+a[\tau^8c_8+\tau^{12}c_6^2/2])+o(a)\,.
\end{equation}
Multiplying this expression on the pre-exponential factor
$$
\sum_{m=3}^7a^{\frac{m-3} 4}\tau^m
(b_m)=\tau^3b_3+a^{1/4}\tau^4b_4+a^{1/2}
\tau^5b_5+a^{3/4}\tau^6b_6+a\tau^7b_7
$$
and contracting with $\dot x^\nu$ we get
\begin{equation}
\begin{array}{lll}
&&e^{-\tau^4/12}\dot{x}^\nu(\tau^3b_3+  \\
&&  \\
&&+a^{1/4}\tau^4b_4+  \\
&&  \\
&&+a^{1/2}[\tau^5b_5+b_3c_6\tau^9]+ \\
&& \\
&&+a^{3/4}[\tau^6b_6+(c_6b_4+c_7b_3)\tau^{10}]+   \\
&&  \nonumber \\
&&+a[\tau^7b_7+(c_6b_5+c_7b_4+c_8b_3)\tau^{11}+\frac
12c_6^2b_3\tau^{15}])_{\nu\mu}+o(a)\,.
\end{array}
\end{equation}
Substituting the last expression into (\ref{I}) and using the
integration formula
\begin{equation}
\int\limits_0^{\infty}e^{-\tau^4/12}\tau^n
d\tau=\frac{12^{\frac{n+1}4}}{4}\Gamma \left( \frac{n+1}4\right)
\end{equation}
we finally arrive at (\ref{rrf}). It is interesting to note that
the resulting expression for $F^{rr}_\mu(s,a)$ turns out to be
orthogonal to both $\dot x^\mu$ and $\ddot x^\mu$.

\section*{Acknowledgments}

We thank V. G. Bagrov and S. L. Lyakhovich for fruitful
discussions. The work was partially supported by RFBR under the
grant no 00-02-17-956, INTAS under the grant no 00-262 and Russian
Ministry of Education under the grant E-00-3.3-184. The work of
AAS was supported by RFBR grant for support of young scientists
no 02-02-06879.

\end{document}